\documentclass[12pt]{article}
\usepackage{amsfonts}

\newcommand{\be}{\begin{equation}}
\newcommand{\ee}{\end{equation}}
\newcommand{\beqa}{\begin{eqnarray}}
\newcommand{\eeqa}{\end{eqnarray}}
\newcommand{\bean}{\begin{eqnarray*}}
\newcommand{\eean}{\end{eqnarray*}}
\newcommand{\eqn}[1]{(\ref{#1})}
\newcommand{\nn}{\nonumber}

\def\sqr#1#2{{\vcenter{\vbox{\hrule height.#2pt
        \hbox{\vrule width.#2pt height#1pt \kern#1pt
                        \vrule width.#2pt}
        \hrule height.#2pt}}}}
\def\square{\mathop\sqr68}

\begin{document}
\title{
Nonlinear gravitational waves and their polarization.}
\author{F. Canfora, G. Vilasi and P. Vitale \\
Dipartimento di Fisica ``E.R.Caianiello'', Universit\`{a} di Salerno.\\
 Istituto Nazionale di Fisica Nucleare, GC di Salerno, Italy. }
\date{today}
\maketitle
\begin{abstract}
Vacuum gravitational fields invariant for a non Abelian Lie algebra
generated by two Killing fields whose commutator is {\it light-like} are
analyzed. It is shown that they represent nonlinear gravitational waves
obeying to two nonlinear superposition laws. The energy and the
polarization of this family of waves are explicitely evaluated.


\end{abstract}

In the last years, together with the experimental efforts devoted to the
detection of gravitational waves, there is a strong theoretical activity to
describe and predict the emission of gravitational waves from astrophysical
systems in strong field conditions. However, all the experimental devices,
laser interferometers (Geo-600, Ligo, Virgo, Tama-300) and resonant
antennas (Allegro, Auriga, Explorer, Nautilus, Niobe), are constructed
according to results obtained by the linearized Einstein field equations,
in close analogy with what is normally done in Maxwell theory of
electromagnetic fields. Indeed, since the sources are at enormous distance
from the Earth, the amplitude of the waves when they reach the detector is
so small that it has always been assumed that when treating the waves in
the Earth's neighborhood the linearized theory suffices. This assumption is
misleading. The nonlinearity of Einstein's equations manifests itself in a
permanent displacement of the test masses of a laser interferometer
detector after the passage of a wave train which cannot be neglected.
Indeed, it has been shown \cite{Ch91} that every gravitational-wave burst
has a nonlinear memory due to the cumulative contribution of the effective
stress of the gravitational waves themselves and that, in the case of a
binary coalescence, the nonlinear memory is of the same order of magnitude
as the maximal amplitude of the dynamical part of the burst.

On the other hand, the non linearity of the gravitational field is one of
its most characteristic properties, and it is likely that at least some of
the crucial properties of the field show themselves only through the
nonlinear terms. Thus, a great deal of interest is still devoted to exact
solutions which more easily enable to discriminate between physical and
pathological features.

Starting from the seventy's new powerful mathematical methods have been
developed to deal with nonlinear evolution equations and their exact
solutions. Indeed, it has been shown \cite{BZ78} that the vacuum Einstein
field equations for a metric of the form
\[
g=f\left( z,t\right) \left( dt^{2}-dz^{2}\right) +h_{11}\left( z,t\right)
dx^{2}+ h_{22}\left( z,t\right) dy^{2}+2h_{12}\left( z,t\right) dxdy
\]
can be integrated by using a suitable generalization of the {\it Inverse
Scattering Transform }\cite{BZ78}, leading to {\it solitary wave
solutions.}

A geometric inspection of the metric above shows that it is invariant under
translations along the $x,y$-axes, {\it i.e. }it admits two Killing fields
$\partial_x$ and $\partial_y$ closing a Abelian $2$
-dimensional Lie algebra\footnote{
The study of metrics invariant for a Abelian $2$-dimensional Killing Lie
algebra goes back to Einstein, Rosen \cite{ER37,Ro54} and Kompaneyets \cite
{Ko58}; recent analysis can be found in \cite{Ve93,NN01}.} ${\cal A}_{2}$.
Moreover, the distribution ${\cal D}$, generated by $\partial_x$ and
$\partial_y$, is $2$ -dimensional and the distribution ${\cal D}^{{\cal
\perp }}$, orthogonal to ${\cal D}$, is integrable and transversal to
${\cal D}$.

Thus, it has been natural to consider \cite{SVV00} the general problem of
characterizing all gravitational fields $g$ invariant for a Lie algebra $
{\cal G}$ of Killing fields such that:

{\bf I.} the distribution ${\cal D}$ generated by the vector fields of
${\cal \ G}$ is $2$-dimensional.

{\bf II.} the distribution ${\cal D}^{{\cal \perp }}$ orthogonal to ${\cal
D} $ is integrable and transversal to ${\cal D}$.

The aim of this letter is to study the physical properties of a subfamily
of such Einstein metrics, better specified below, which has the important
characteristic of representing propagative gravitational fields. The wave
character of these solutions will be established on a rigorous basis and
the problem of defining physical observables like the energy flux and the
polarization will be addressed.

In the following  a sub-algebra of the Lie algebra of all Killing fields of
a metric $g$ will be called {\it Killing algebra} of $g$.  Moreover, an
integral ($2$-dimensional) submanifold of ${\cal D}$ will be called a {\it
Killing leaf}.

\paragraph*{Invariant metrics.} If $g$ is a metric on the space-time and
${\cal G}_{2}$ one of its Killing algebras whose generators $X,Y$ satisfy
\[
\lbrack X,Y]=Y,\ \ \
\]
then, under the conditions I and II, a coordinate system $(x^{\mu })$, $\mu
=1,2,3,4$, adapted to the Killing fields, exists \cite{SVV00} such
that
\[
X=\partial_3 ,\,\,\ \,Y=e^{x^{3}}\partial_4,
\]
and the most general ${\cal G}_{2}$-invariant metric has the form
\begin{eqnarray*}
g &=&g_{ij}dx^{i}dx^{j}+\left( \lambda \left( x^{4}\right)^{2}-2\mu
x^{4}+\nu \right) \left(dx^{3}\right)^2+ \\ &&2\left( \mu -\lambda
x^{4}\right) dx^{3}dx^{4}+\lambda \left(dx^{4}\right)^2,\quad i,j=1,2
\end{eqnarray*} where\ $g_{ij}$, $\lambda $, $\mu $, and $\nu $ are
arbitrary functions of $\left( x^{1},x^{2}\right) $.

\paragraph*{Einstein metrics.} If the Killing field $Y$ is of {\it light type}, {\it
i.e}. $g(Y,Y)=0$,  then the general solution of vacuum Einstein equations,
in the adapted coordinates $\left(x^{1},x^{2},x^{3},x^{4}\right) $, is
given \cite{SVV00} by
\be
g=2f\left(\left(dx^{1}\right)^{2}+\left(dx^{2}\right)^{2}\right)
+\mu\left[(w\left( x^{1},x^{2}\right)
-2x^{4})\left(dx^{3}\right)^{2}
+ 2dx^{3}dx^{4}\right],  \label{gm}
\ee
where $\mu =D\Phi +B$; $D,B\in {\cal R}$, $\Phi $ is a non constant
harmonic function of $(x^1, x^2)$, $f=\left( \nabla \Phi \right)^{2}
\sqrt{\left|\mu\right| }/\mu $, and $w$ is a solution of the
$\mu $-{\it deformed Laplace equation:}
\begin{equation}
\Delta w+\left( \partial_{1}\ln \left| \mu \right| \right) \partial_{1}w
+\left( \partial_{2}\ln \left| \mu \right| \right) \partial_{2} w=0.
\label{dle}
\end{equation}
Solutions of the $\mu $-deformed Laplace equation will be called
 $\mu $-{\it harmonic functions}.

Despite the non-linear nature of general relativity, it is possible to
exhibit two {\it superposition laws} for the gravitational fields
(\ref{gm}). Indeed, with {\it two} harmonic functions $\Phi _{1}$ and $\Phi
_{2}$ we can associate {\it three} gravitational fields (in facts a whole
2-parameters family), that is, $g_{\Phi _{1}}$, $g_{\Phi _{2}}$ and
$g_{a\Phi _{1}+b\Phi _{2}}$; the last one, which is associated with a
linear combination of $\Phi _{1}$ and $\Phi _{2}$, is to be intended as the
superposition of the two associated solutions $g_{\Phi _{1}}$ and $g_{\Phi
_{2}}$. The second superposition law follows from the linearity of the $\mu
$-deformed Laplace equation, so that with two $\mu $-harmonic functions
$w_{1}$ and $w_{2}$ we can associate the gravitational fields $g_{w_{1}}$,
$g_{w_{2}}$ and $g_{aw_{1}+bw_{2}}$.

The coordinates $(x^{3},x^{4})$ on the Killing leaves have a clear
geometric meaning but are of difficult physical interpretation.
Fortunately, being the Killing leaves flat manifolds \cite{SVV00} it is
possible to adopt coordinates diagonalizing the metric restricted to the
Killing leaves.

This allows to define {\it harmonic coordinates} $\left( x,y,z,t\right)$
  such that the metrics (\ref{gm}) take a simpler and physically more
interesting form \cite{CVV02}. If, for the sake of simplicity, we choose
$\mu =1$, then
\[
\left\{
\begin{array}{l}
x=x^{1} \\ y=x^{2} \\ z=\frac{1}{2}\left[ \left( 2x^{4}-w\left(
x^{1},x^{2}\right) \right) \exp
\left( -x^{3}\right) +\exp \left( x^{3}\right) \right]  \\
t=\frac{1}{2}\left[ \left( 2x^{4}-w\left( x^{1},x^{2}\right) \right) \exp
\left( -x^{3}\right) -\exp \left( x^{3}\right) \right] .
\end{array}
\right.
\]
In these coordinates, the Einstein metrics (\ref{gm}) (with $\mu =1$) read
\be
g=2\left( \nabla \Phi \right)^{2}[dx^{2}+dy^{2}] +dz^{2}-dt^{2}+dw~d\ln
\left|z-t\right|,  \label{ggen}
\ee
where $t$ plays the role of {\it time}.

From the physical point of view it is already clear that the Einstein
metrics (\ref{ggen}) represent {\it gravitational waves}, the {\it
wave-like }components, in the almost Minkowskian coordinates $\left(
x,y,z,t\right) $, slowly decreasing when $\left| z-t\right| \rightarrow
\infty $.

The choice of harmonic coordinates makes it clear that:
\begin{itemize}
\item  when $w$ is constant the Einstein metrics (\ref{ggen}) admit the
time-like Killing vector field $\partial_t$, which is orthogonal to
space-like hypersurfaces defined by $t=$ {\it const}, and then represent
{\it static} gravitational fields. Under the further assumption $\Phi
=x{/\sqrt{2}}$, they reduce to the Minkowski metric.
\item  when $w$ is not constant the Einstein metrics (\ref{ggen}) represent
a {\it disturb} moving along the$\ z$ direction on the Killing leaves at
light velocity.
\end{itemize}

In the following we will assume that $w$ is not constant. The choice $\mu
=1$ allows to simplify the computational algebra without spoiling the most
interesting features of the metrics like the superposition law and the
propagative character. Moreover, the form (\ref{ggen}) will result
particularly manageable for a discussion about the definition of spin and
energy. Explicit results will be illustrated mainly for the case $ f= ${\it
const}.

\paragraph*{Wave character of the gravitational field.}
The second step to attempt a
physical interpretation of the solutions we are considering is the study of
their wave character.

The gravitational fields (\ref{ggen}) satisfy  the Zakharov generalization
of the Zel'manov criterion \cite{Za73} which states that a vacuum solution
of the Einstein equations is a gravitational wave if the corresponding
Riemann tensor field $R$ is not covariantly constant and its components
$R_{\mu \nu \lambda \sigma }$ satisfy the hyperbolic equation
\[
g^{\alpha \beta }\nabla _{\alpha }\nabla _{\beta }R_{\mu \nu \lambda \sigma
}=K_{\mu \nu \lambda \sigma },
\]
where $\ \nabla _{\beta }$ denotes the Levi-Civita covariant derivative and
$K$ is a tensor field depending at most on first derivatives of the Riemann
tensor field. The criterion is certainly satisfied if
\begin{equation}
g^{\alpha \beta }\partial_{\alpha }\partial_{\beta }R_{\mu \nu \lambda
\sigma }=0,  \label{zel}
\end{equation}
where $\partial_{\alpha }$ are now ordinary partial derivatives.

Concerning the physical meaning of this criterion, it can be shown
\cite{Za73} that the characteristic hypersurface of the system of equations
(\ref{zel}) is identical with the characteristic hypersurface of the
Einstein and Maxwell equations in curved space-time. Consequently
Eqs.(\ref{zel}) describe the propagation of the discontinuities of the
second derivatives of the Riemann tensor. This links the Zel'manov
criterion to the intuitive concept of {\it local wave of curvature}.

In the simple case when the function $f$ is constant the criterion is
easily verified. In facts, the only independent non vanishing components of
the Riemann tensor field  are
\be
R_{txzx}=\frac{{\partial_x}^2 w}{2(z-t)^{2}},~~~ R_{txzy}
=\frac{\partial_x\partial_y w}{ 2(z-t)^{2}},~~~
 R_{tyzy}=\frac{{\partial_y}^2 w}{2(z-t)^{2}};
\label{riemann}
\ee
since $w$ is a   harmonic function in the plane $(x,y)$, they
 automatically satisfy the condition (\ref{zel}), the derivatives of
$w$ being harmonic functions as well. With a little computational effort
\cite{CVV02} the result generalizes to $ f,\mu $ generic. Thus, the wave
character of the gravitational fields (\ref {ggen}) is established
according to a rigorous and intrinsic criterion.

\paragraph*{The energy-momentum pseudo-tensor.} For$\,\,f=1/2$, the exact solution
(\ref{ggen}) reduces to
\begin{equation}
g=dx^{2}+dy^{2}+dz^{2}-dt^{2}+dwd\ln \left| z-t\right|  \label{gw}
\end{equation}
and has the physically interesting form of a {\it perturbed Minkowski
metric}, $g=\eta +h$, with $\eta $ the Minkowski metric and
\beqa
h&=&dw~d\ln \left| z-t\right|=\frac{1}{z-t}\left[ (\partial_x w) dx
dz\right. \nn\\ &+&\left.(\partial_y w) dy dz-(\partial_x w) dx
dt-(\partial_y w) dy dt\right]
\label{traceless}
\eeqa
 It is easily verified that the metric
\eqn{gw}, besides being a exact solution of the vacuum Einstein equations,
satisfies the linearized equations ${\square} h_{\mu\nu}=0$ as well.

To study the energy content and the polarization of gravitational waves it
is convenient to use the {\it Landau-Lifshitz} {\it energy-momentum
pseudo-tensor} \cite{LL65} which has been seen to yield the correct
definition of energy for relevant cases \cite{PP79}. In facts, the energy
flux radiated at infinity for an asymptotically flat space-time, evaluated
with the Landau-Lifshitz pseudo-tensor, has been seen to agree with the
Bondi flux \cite{BVM62} that is with the energy flux evaluated in the exact
theory. For gravitational fields (\ref{gw}) it reduces to
\begin{eqnarray}
\tau ^{\rho \kappa } &=&C\left\{ -\Gamma _{\lambda \sigma }^{\nu }\Gamma
_{\mu \nu }^{\sigma }(g^{\rho \lambda }g^{\kappa \mu }-g^{\rho \kappa
}g^{\lambda \mu })\right.  \nonumber \\ &-& \left. g^{\rho \lambda }g^{\mu
\nu }\Gamma _{\nu \sigma }^{\kappa }\Gamma
_{\lambda \mu }^{\sigma }-g^{\kappa \lambda }g^{\mu \nu }\Gamma _{\nu \sigma
}^{\rho }\Gamma _{\lambda \mu }^{\sigma }  +  g^{\mu
\lambda }g^{\sigma \nu }\Gamma _{\lambda \nu }^{\rho
}\Gamma _{\mu \sigma }^{\kappa }\right\}  \label{landauhar}
\end{eqnarray}
where the harmonicity condition $\Gamma _{\mu \nu }^{\lambda }g^{\mu \nu
}=0$ and the property $\eta ^{\mu \nu }h_{\mu \nu }=0$, which in turn
implies $\Gamma _{\nu \sigma }^{\sigma }=0$, have been used.

Relevant are the components $p^{\mu }\equiv \tau _{0}^{\mu }$ of the mixed
pseudo-tensor, $p^{\mu }$ being the $4$-momentum density,
\begin{eqnarray*}
p^{0} &=&f^{-2}\left( t-z\right)^{-2}[C_{1}\left({\partial_x}^2
w\right)^{2}+C_{2}\left(\partial_x\partial_y w\right)^{2}]
\\ &+&f^{-2}\left(t-z\right) ^{-4}C_{3}\nabla\cdot
\left(\left|\nabla w\right|^2
\nabla w\right) \\
 p^{1} &=&p^{2}=0 \\ p^{3}&=&p^{0}
\end{eqnarray*}
where $C_{i}$ are positive numerical constants and the harmonicity
condition for $w$ has been used.

Concerning the definition of the polarization, the above form for
$\tau_{0}^{\mu }$ is particularly appealing because, apart from a
physically irrelevant total derivative, the component $\tau _{0}^{0}$,
representing the energy density, is expressed as a sum of square
amplitudes. Moreover, the momentum $p^{i}=\tau _{0}^{i}$ is only non
vanishing in the direction of the propagation of the wave and it is
proportional to the energy with proportionality constant $c=1$ which is the
light velocity in natural units.

The study of the polarization of this wave leads to a non standard result,
that is $h_{tx}=\left( z-t\right) ^{-1}\partial_{x}w$ and $h_{ty}=\left(
z-t\right) ^{-1}\partial_{y}w$ {\it represent a spin one field}. This is a
consequence of the fact that we are not in the transverse-traceless gauge
or, roughly speaking, $h$ has only one index in the plane transverse to the
propagation direction.

\paragraph*{The polarization.} The definition
and the meaning of spin or polarization
for a theory, such as general relativity, which is non-linear and possesses
a much bigger invariance than just the Poincar\'{e} one, deserve a careful
analysis.

It is well known that the concept of particle together with its degrees of
freedom like the spin may be only introduced for linear theories (for
example for the Yang-Mills theories, which are non linear, we need to
perform a perturbative expansion around the linearized theory). In these
theories, when Poincar\'{e} invariant, the particles are classified in terms of
the eigenvalues of the two Casimir operators of the Poincar\'{e} group, $P^{2}
$ and $W^{2}$ where $P_{\mu }$ are the translation generators and $W_{\mu
}={\frac{1}{2}}\epsilon _{\mu \nu \rho \sigma }P^{\nu }M^{\rho \sigma }$ is
the Pauli-Ljubanski polarization vector with $M^{\mu \nu }$ Lorentz
generators. Then, the total angular momentum $J=L+S$ is defined in terms of
the generators $M_{\mu \nu }$ as
$J^{i}={\frac{1}{2}}\epsilon^{0ijk}M_{jk}$. The generators $P_{\mu }$ and
$M_{\mu \nu }$ span the Poincar\'{e} algebra, ${\cal ISO}(3,1)$.

When $P^{2}=0$, $W^{2}=0$, $W$ and $P$ are linearly dependent
\[
W_{\mu }=\lambda P_{\mu };
\]
the constant of proportionality is the {\it helicity}. The time component
of $W$ is $W^{0}=\overrightarrow{P}\cdot \overrightarrow{J}$, so that
\[
\lambda =\frac{\overrightarrow{P}\cdot \overrightarrow{J}}{P_{0}},
\]
which is the definition of helicity for massless particles like the
photons.

Let us turn now to the gravitational fields represented by Eq. (\ref{gw}).
As it has been shown, they represent gravitational waves moving at the
velocity of light, that is, in the would be quantized theory, particles
with zero rest mass. Thus, if a classification in terms of Poincar\'{e} group
invariants could be performed, these waves would belong to the class of
unitary (infinite-dimensional) representations of the Poincar\'{e} group
characterized by $P^{2}=0$, $W^{2}=0$. In order for such a classification
to be meaningful $P^{2}$ and $W^{2}$ have to be invariants of the theory.
This is not the case for general relativity, unless we restrict to a subset
of transformations selected for example by some physical criterion or by
experimental constraints. For the solutions of the linearized vacuum
Einstein equations the Lorentz transformations are selected out by the
request that the energy--momentum pseudo-tensor be truely a tensor and that
the direction of propagation of the waves be preserved  \cite{Di75}.

Let us see what happens in our case. There exist several equivalent
procedures to evaluate the polarization. One of these consists in looking
at the $\tau _{0}^{0}$ component of the Landau-Lifshitz pseudo-tensor, and
to see how the metric components that appear in $\tau _{0}^{0}$  transform
under an infinitesimal rotation ${\cal R}$ in the plane $\left( x,y\right)
$ transverse to the propagation direction\footnote{With respect to the
Minkowskian background metric this plane is orthogonal to the propagation
direction. With respect to the full metric it is transversal to the
propagation direction and orthogonal only in the limit $
\left| z-t\right| \mapsto \infty$.}. This transformation preserves the
harmonicity condition. The physical components of the metric are $h_{tx}$
and $h_{ty}$ and under the infinitesimal rotation ${\cal R}$ in the plane
$\left( x,y\right) $ they transform as components of a vector. Applied to
any vector $\left( v^{1},v^{2}\right) $, the infinitesimal rotation ${\cal
R}$ acts as
\[
{\cal R}v^{1}=v^{2}\,,\,\,\,\,\,{\cal R}v^{2}=-v^{1}.
\]
Then,
\[
{\cal R}^{2}v^{i}=-v^{i}\,\,\,\,\ i=1,2,
\]
so that $i{\cal R}$ has eigenvalues $\pm 1$. Thus, the components of
$h_{\mu\nu }$ which contribute to the energy density correspond to spin $1$
fields. The reason for this is that the propagating part of the metric, in
the harmonic coordinates, has a off-diagonal form or, put in another way,
$h_{\mu \nu }$ has one index in the propagation direction and one in the
orthogonal plane. As in the linearized theory, of the whole diffeomorphisms
group just the Lorentz transformations preserve the propagation direction
and ensure the tensorial character of the energy--momentum pseudo--tensor.
Moreover, they preserve the harmonic gauge.

The observable effects of the gravitational waves (\ref{gw}) follow from
the study of the relative motion of test particles described by the {\it
geodesic deviation equation}. Since the only nonvanishing components of the
Riemann tensor field $R_{\mu \alpha \nu }^{\lambda }$ are listed in
\eqn{riemann}, for small velocity  and in the
weak field approximation the geodesic deviation equations reduce to
\[
{\partial_t}^{2} V^{i }=\frac{V^{j } \eta^{ik}}{(z-t)^{2}}
\partial_k\partial_j w,
~~~{\rm for}~ i=1,2,3
\]
the vector field $V\equiv(0,V^1, V^2,V^3)$ representing the {\it
space-like} separation of close geodesics. Thus, it is clear that the
deviation depends explicitly on the choice of the harmonic function $w$. A
complete analysis will be performed in a forthcoming paper by introducing a
source for the gravitational waves here described.


\begin{thebibliography}{99}
\bibitem{Ch91}  D. Christodoulou, {Phys. Rev. Lett.} {\bf 67}, 1486
(1991).
\bibitem{BZ78}  V. A. Belinsky and V. E. Zakharov, {Sov. Phys. Jetp} {\bf 48},
 6 (1978); {\bf 50}, 1 (1979).
\bibitem{ER37}  A. Einstein and N. Rosen, {J.Franklin Inst.} {\bf 223}, 43
(1937).
\bibitem{Ro54}  N. Rosen, {Bull. Res. Coun. Isr.} {\bf 3}, 328 (1954).
\bibitem{Ko58}  A. S. Kompaneyets, {Sov. Phys. JETP} {\bf 7}, 659 (1958).
\bibitem{Ve93}  E. Verdaguer, {Phys. Rep.} {\bf 229},1 (1993).
\bibitem{NN01}  H. Nicolai and A. Nagar,  in {\it Gravitational Waves} IOP,
         245 (2001) (I. Ciufolini et al. eds.).
\bibitem{SVV00}  G. Sparano, G. Vilasi and A. M. Vinogradov, {Phys. Lett.}
{\bf B 513}, 142 (2001); {Diff. Geom. Appl.} {\bf 16}, 95 (2002).
\bibitem{CVV02}  F. Canfora, G. Vilasi and P. Vitale, in preparation.
\bibitem{Za73}  V. D. Zakharov, {\it Gravitational waves in Einstein's
theory,} (Halsted Press, New York, 1973).
\bibitem{LL65}  L. D. Landau and E. E. Lifshitz, {\it Theorie du champ,} (Mir,
Moscow, 1965).
\bibitem{PP79}  S. Persides and D. Papadopoulos, {Gen. Rel. Grav.} {\bf
11}, 233 (1979).
\bibitem{BVM62}  H. Bondi, M. G. J. van der Burg and A. W. K. Metzner, {Proc.
Roy. Soc. }{\bf A269}, 21 (1962).
\bibitem{Di75}  P. A. M. Dirac, {\it General Theory of Relativity}
(J. Wiley \& Sons, New York, 1975).
\end{thebibliography}
\end{document}